\documentclass[prf,amsmath,amssymb,groupedaddress,notitlepage,longbibliography]{revtex4-2}

\usepackage{lmodern} 
\usepackage[T1]{fontenc}
\usepackage{textcomp}
\usepackage[scr=boondoxo,frak=boondox,bb=boondox]{mathalfa}

\usepackage[colorlinks=true, linkcolor=Blue, allcolors=Blue]{hyperref}

\usepackage{fancyhdr} 
\usepackage{graphicx}
\usepackage[usenames, dvipsnames]{color}
\usepackage[export]{adjustbox}

\newcommand{\RNum}[1]{\uppercase\expandafter{\romannumeral #1\relax}}

\usepackage[squaren, Gray, cdot]{SIunits}

\begin{document}

\title{Lift-up and streak waviness drive the self-sustained process \\ in wall-bounded transition to turbulence}  

\author{Tao Liu\textsuperscript{1}}
\author{Beno\^{i}t Semin\textsuperscript{1}}
\email{benoit.semin@espci.fr}
\author{Ramiro Godoy-Diana\textsuperscript{1}}
\author{Jos\'e Eduardo Wesfreid\textsuperscript{1}}
\affiliation{\textsuperscript{1} PMMH, CNRS UMR 7636, ESPCI Paris---Universit\'e PSL, Sorbonne Universit\'e, Universit\'e  Paris Cit\'e, F-75005 Paris, France}

\begin{abstract}
Flow field measurements from a Couette-Poiseuille experiment are used to examine quantitatively certain steps of  the self-sustained process (SSP) of wall-bounded transition to turbulence. Although the different parts of the SSP have been discussed at large in the literature, direct measurements from experiment are scarce and, to our knowledge, the present results are the first to show, using a local analysis of the turbulent patterns, that: (1) the amplitude of streamwise rolls is related to streak waviness, bringing a quantitative picture to one of the main physical mechanisms of Waleffe's model of SSP ; and (2), at low waviness, direct measurements of the correlation between the streak and roll amplitudes, respectively probed by the streamwise and wall-normal velocity perturbations, quantify the lift-up effect.   
\end{abstract}
\maketitle
\thispagestyle{empty}

\section{Introduction}

Turbulent structures in the form of streamwise elongated structures (streaks) and streamwise vortices (rolls) are recognised as important features in transition to turbulence in plane shear flows. These structures have a long history of being observed in boundary layer flow experiments \cite{elder_1960,klebanoff_tidstrom_sargent_1962,kline_reynolds_schraub_runstadler_1967}. The dynamics of streaks and rolls in plane shear flows are essential for understanding the process of transition to turbulence and the maintenance of turbulence at high Reynolds number. Turbulence in confined plane shear flows can become self-sustained above a critical Reynolds number $Re_c$ through a regeneration cycle so-called self-sustaining process (SSP) \cite{jimenez_moin_1991,hamilton_kim_waleffe_1995,Waleffe1997SSP}, which describes the interplay between the elongated steaks and rolls forming a closed cycle. 

The elongated streaks are generated by the advection by the rolls in a velocity field with gradients as in presence of a wall-bounded or confined velocity flow,  via the lift-up effect \cite{Landahl1975}. The streaks then experience wave-like instabilities, become unstable and break down, which results in the re-injection of energy into the rolls by non-linear interactions \cite{Waleffe1997SSP}. The basic mechanisms of this process are well described by minimal models that couple the evolution of modes for streaks, rolls, wall-normal vorticity, and mean flow deformation \cite{Waleffe1997SSP,Moehlis:2004,Manneville:2018,Cavalieri:2022}, and by alternative models in terms of vortex-wave interactions \cite{hall_sherwin_2010}. The non-linear dynamics and the phase space of the SSP has been studied in plane Couette flow \cite{Dauchot2000}.

The interaction between streaks and rolls has been investigated lately by stochastic structural stability theory (SSST) to show the forcing from the streaks to the rolls via Reynolds stresses \cite{farrell_ioannou_2012}. The SSP process has also been revealed in Taylor-Couette flow simulations, in which this mechanism is studied by quantifying the energy from different turbulent structure components \cite{Dessup:2018,Sacco:2019,Martinand:2014,Wang:2022}. In a turbulent channel flow simulation, resolvent analysis has been used to determine the main forcing mode (a pair of streamwise rolls) that accommodates and causes the most kinetic energy in the buffer and logarithmic layers \cite{bae_2021}. The non-linear interactions between the rolls and the oblique streaks were found responsible for the self-sustaining turbulence in a minimal channel \cite{bae_2021,nogueira_morra_martini_cavalieri_henningson_2021}. 
The stability and breakdown of large-scale streaks have been studied numerically and experimentally in boundary-layer and channel flows \cite{Mans2005, Mans2007, MansPhdThesis, brandt_schlatter_henningson_2004, Brandt_2008,Cossu:2017}. Mans \cite{MansPhdThesis} reported experimental measurements of a sinuous and a varicose instability of the streaks that can cause roll-up structures after reaching a critical amplitude in free-stream turbulence boundary layer. Using vortex generators in a boundary layer, Duriez~\emph{et al.}\cite{Duriez2009} measured  an energy transfer from the streaks to the rolls via streaks instabilities at moderate Reynolds number, which is an experimental evidence of the SSP process. 

Now, concerning wavy streaks and their interactions, experimental measurements are particularly lacking, especially regarding the role of streak waviness in their regeneration cycle. Numerically, recent interest has risen on the study of local waviness  \cite{Hwang:2022}. In this paper, we investigate quantitatively the link between this waviness and the other velocity components using a well-controlled wall-bounded plane shear flow experiment: a Couette-Poiseuille flow (CPF) channel (see Fig.~\ref{fig:expsetup}), and stereo particle image velocimetry (PIV) measurements. To allow both straight and wavy streaks, we examine Reynolds numbers around the value  $Re_g\approx680$ above which turbulence has been shown to be self-sustained in CPF \citep{Klotz2017PRF, liu_2021}, and also at higher values to be in a fully turbulent regime. Using a local streak analysis approach based on the stereo-PIV velocity field measurements, we investigate the local relation between the waviness of the streaks and the rolls velocity components, bringing a quantitative assessment of different parts of the SSP. The article is organized as follows. We describe the experimental set-up and protocol in section \RNum{2}. The stereo-PIV velocity field analysis and the local streak processing method are described in section \RNum{3}. We investigate the local relation between the waviness of the streaks and the rolls velocity components in section \RNum{4}. The discussion and conclusions are contained in section \RNum{5}.

\section{Experimental set-up and protocol}

\subsection{Couette-Poiseuille channel}

\begin{figure}
	\centerline
	{\includegraphics[width=0.8\linewidth]{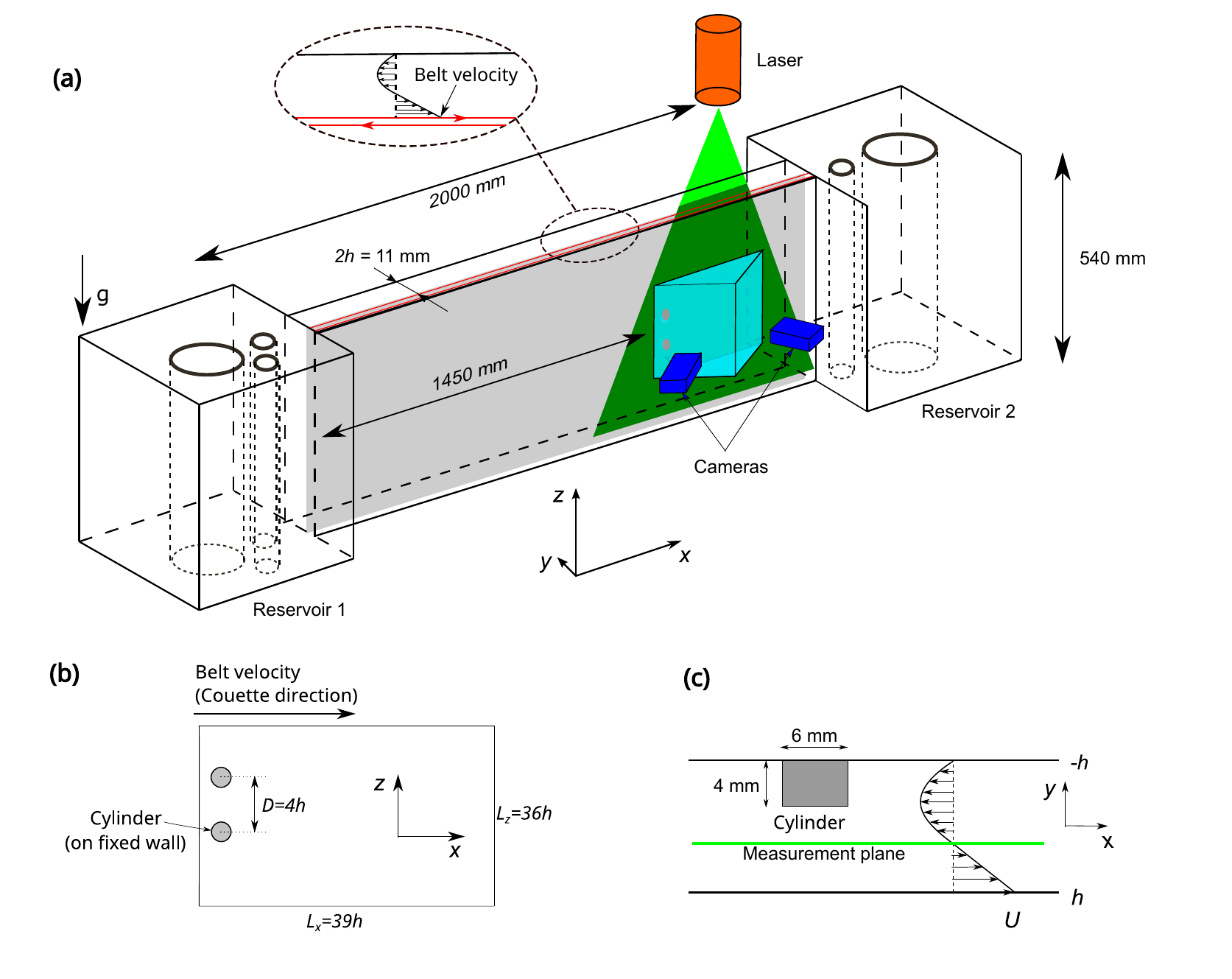}}
	\caption{(a) Schematic diagram of the experimental Couette-Poiseuille set-up and stereo-PIV configuration.  (b) Position of the cylindrical vortex generators on the vertical $xz$ plane. (c) Side view of the vortex generator in the $xy$ plane. The velocity profile displayed is the laminar Couette-Poiseuille.}
	\label{fig:expsetup}
\end{figure}

We use the Couette-Poiseuille experiment shown in Fig.~\ref{fig:expsetup}a that has already been described in previous publications \citep{Klotz2017PRF,liu_2021}. It consists of a water channel made of two parallel vertical glass plates with a constant $14$ mm gap, connected to 2 reservoirs at its ends. The glass plates are closed at the top and bottom by two horizontal transparent plates, forming a closed channel. The tops of the reservoirs are not closed. The set-up is filled with water at room temperature $18^{\circ}$C $\pm 1.5^{\circ}$C, which is controlled by air conditioning. Contrary to the work described in \citep{liu_2021}, no grid is placed at the entrance of the channel to induce perturbations.

The flow is driven by a transparent Mylar\textsuperscript{\textregistered} sheet of thickness $\unit{175}{\micro\meter}$. The sheet is closed to form a belt by using double sided tape. The belt is guided by vertical cylinders so that it remains parallel to the glass plates, and close to one of the two plates. A motorised  $\unit{140}{\milli\meter}$-wide cylinder in reservoir~1 drives the belt motion. The speed of the motor (Yaskawa Electrics 100 W servo-motor, gear reduction $1:26$) is set by a controller via a National Instruments card, which is integrated in a Labview\textsuperscript{\textregistered} program. The belt velocity $U_{\rm belt}$ is constant during an experiment, giving a permanent regime for the CPF that is established in the region between the moving belt and the fixed wall. The width of the CPF channel is defined in the usual way as $2h$, with $h=5.5$~mm being the half channel width. In the laminar regime, a parabolic Couette-Poiseuille profile is obtained. The length of the channel in the streamwise direction, i.e. the $x$ direction, is $L_x = 2000$~mm, so that $L_x/h = 364$. The height of the channel in the spanwise direction, i.e. the $z$ direction, is $L_z = 540$~mm, so that $L_z/h = 98$. This aspect ratio is sufficient to investigate the spatio-temporal dynamics of turbulent structures.

The Reynolds number of the Couette-Poiseuille flow is defined as $Re=U_{\rm belt}h/\nu$, where $\nu$ is the kinematic viscosity of the water (evaluated at the measured water temperature). In the following, the variables will be made dimensionless using $h$ and $U_{\rm belt}$.

\subsection{Cylindrical vortex generators}

In order to study the waviness of streaks at Reynolds numbers below the self-sustained threshold ($Re_g\approx 680$ \citep{liu_2021}), two cylindrical vortex generators mounted on the fixed wall inside the channel are used (see Fig.~\ref{fig:expsetup}b and c). This is inspired by previous works in boundary layers that obtained experimental evidence of the self-sustaining process \cite{Duriez2009} and secondary flow instabilities \cite{fransson2005,Cossu2010JOT}. In a boundary layer, each wall-mounted vortex generator induces a pair of longitudinal counter-rotating vortices that modulate the streamwise velocity profile close to the wall by the lift-up effect (see \cite{Brandt:2014_LiftupEffectLinear} for a review). In the channel of the present experiments, the flow induced by the cylinders is probably more complex close to the cylinders, due to the second wall and the non-monotonous Couette-Poiseuille base flow profile. However, a couple of cylinders were still found to be an efficient method to induce streaks.

The cylinders are aligned in the $z$ direction with a spacing of $D = 4h$, which is close to the optimal  distance between streaks  \cite{Klotz:2017_transient_growth} (see Fig.~\ref{fig:expsetup}b). The height and the diameter of the cylinder are $ 4$~mm and $6$~mm,  respectively. The chosen height is sufficient to generate quasi-streamwise streaks at low $Re$ (ex: $Re<500$) outside the  very near region to the bluff bodies ($x < 5h$). Cylinders with a height of $2$ mm were not able to induce streaks at low $Re$. The cylinders used are actually Neodymium disc magnets (from Supermagnete\textsuperscript{\textregistered}) so that they can easily be held and manipulated on the fixed wall.

\subsection{Stereo-PIV measurements}

Stereo particle image velocimetry (PIV) is used to measure the three velocity components in a 2D streamwise-spanwise $xz$ plane parallel to the walls of the channel, at a distance of  $0.33h$ from the moving wall, where the laminar base flow vanishes ---see Fig.~\ref{fig:expsetup} (a) and (c). An integrated stereo-PIV system from LaVision\textsuperscript{\textregistered} is used, with a Darwin-Duo\textsuperscript{\textregistered} $20$~mJ Nd-YLF double pulse green laser ($527$~nm) to generate the laser sheet. The laser sheet is aligned with the channel using a rotation and tilting stage (Standa 6PT110), and the position of the sheet in the channel is selected using a Thorlabs\textsuperscript{\textregistered} micro-translation stage.
The flow is seeded with $\unit{20}{\micro\meter}$ polyamid beads of density $\unit{1.03}{\gram \, \centi\meter^{-3}}$, with a volumic concentration equal to $\unit{1.7\times 10^{-5}}{\gram \,\centi\meter^{-3}}$.

The images are acquired by two double-frame cameras Imager MX5M\textsuperscript{\textregistered} from LaVision ($2664 \times 2056$~pixels), each one equipped with a Nikon\textsuperscript{\textregistered} $17-35$~mm objective lens set at an aperture f/2.8. The cameras were mounted at a distance of about 770~mm from the measurement plane. The angle between the cameras was approximatively of $\unit{90}{\degree}$. A water tank of angle $\unit{45}{\degree}$ is placed between the camera and the channel to reduce optical distortions. A Scheimpflug adapter is used to tilt the camera so that the whole object plane is in focus. The camera, the Scheimpflug adapter, and the objective are mounted on a rotating plate to enable the setting of the angle between the image plane and the camera.
The calibration is achieved by taking two images of a calibration plate, one from each camera. We used a LaVision 2-level double-sided calibration plate (type 058-5) with dimensions: $\unit{58}{\milli\meter} \times \unit{58}{\milli\meter}$, distance between dots 5mm, dot diameter 1.2~mm and level separation 1~mm. The self-calibration process from LaVision software is applied on the PIV particle images to decrease the disparity between the calibration target plane and the laser sheet plane.

The image rate of the acquisition was set at $f^*=8$~Hz, which enables capturing the dynamic evolution of the streaks. The measured field is in the center of the channel. Its size is $29h\times19h$, which gives a good compromise of a small enough PIV interrogation window size of 2~mm (0.36 $h$) to resolve the spatial structures and a large enough global view of these structures. The time interval between the two laser pulses was $\Delta t^* = 4.0$ ms, which gives a particle displacement between two frames of a few pixels. 
The velocity fields are computed using DaVis 10 software (LaVision) with a two-pass algorithm. As the velocity field is dominated by the streamwise velocity component, the displacement of the particles in this direction was more than one order of magnitude larger than in the spanwise direction. For this reason, an elliptical interrogation window with a 4:1 ratio was used, with an area of $1024$ pixels. The overlap between two successive windows was set to $50\%$. 
The 2D velocity field measured from each camera is calculated by cross-correlation and then used for the reconstruction of the three velocity components in the plane based on the calibration of the two cameras.

\section{Velocity field and local streak analysis}

\subsection{Small-scale flow}
\label{Velocity field}

\begin{figure}
	\centering
	\includegraphics[width=0.9\linewidth]{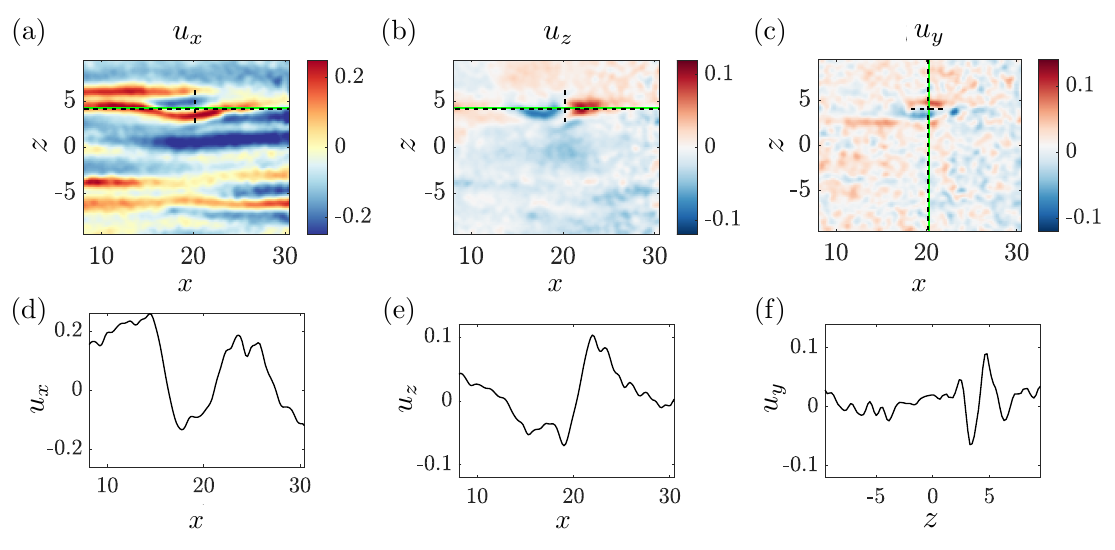}
	\caption{Stereo-PIV velocity fields on the $xz$ plane at $y=0.33h$: (a) streamwise velocity $u_x$, (b) spanwise velocity $u_z$, (c) wall-normal velocity $u_y$. The velocities are non-dimensionalized by the belt velocity $U_{\rm belt}$; (d), (e) and (f) show the velocity profile along the green solid lines in (a), (b), and (c), respectively. These lines are located close to a wavy streak.} 	
	\label{fig:uxuzuyquiver}
\end{figure} 

For studying the streak dynamics in the CPF turbulent patterns, we decompose the measured velocity field $\mathbf{U}$ as the sum of a large-scale flow $\mathbf{U}_{LSF}$ and small-scale flow $\mathbf{u}=(u_x,u_y,u_z)$, i.e.
 \begin{equation}
 \mathbf{U}=\mathbf{U}_{LSF}+\mathbf{u} .
 \label{LSF}
 \end{equation}
 
Large-scale flows originate from the inhomogeneity observed at the interfaces of laminar-turbulent regions \cite{Duguet:2013,Lemoult:2014,Couliou:2015,klotz_pavlenko_wesfreid_2021,Gome:2023_PatternsTransitionalShear}
and their characteristic length scales can be much larger than the typical distance between streaks (see e.g. \cite{Gome:2023_PatternsTransitionalShear}). In experiments, they have also been associated with the geometrical imperfections on the channel walls that induce 3D flows. In the present CPF setup, they have been investigated extensively in a previous work \cite{klotz_pavlenko_wesfreid_2021}. The focus here will be on the local properties of the small-scale turbulent structures described by $u$ in Eq.~\ref{LSF}. In practice, we apply a high-pass circular fourth-order Butterworth spatial filter to separate the flow scales in the 2D $x-z$ plane, with a cut-off wavelength $k_c=0.86/ h$, i.e. $\lambda_c= 7.3 h$ (see \cite{liu_2021,klotz_pavlenko_wesfreid_2021} for details on the choice of $k_c$). 
 
An example of  the streamwise $u_x$, spanwise $u_z$, and wall normal $u_y$ components of the small-scale velocity field measured at $Re=550$ is shown in Fig.~\ref{fig:uxuzuyquiver}. As is always observed in this type of striped three-dimensional unstable flows, the streamwise $u_x$ component is greater than the transverse components $u_y$ and $u_z$. We recall that this patterned structure is observed here even for Reynolds numbers corresponding to laminar and stable situations because of the vortex generators, which induce the streamwise streaks and rolls.
The relationship between the magnitude of the longitudinal fluctuation $u_x$ and that of the wall-normal $u_y$ fluctuation is the basis of the description of the lift-up mechanism, which characterizes the specificity of the transition to turbulence in wall-bounded shear flows, where the streamwise vortices or rolls induce streaks. This relationship will be discussed further below, but a first observation of Fig.~\ref{fig:uxuzuyquiver} (a), (b) and (c) shows that when the streaks visible in the $u_x$ field are relatively straight ---see e.g. the two high speed streaks around $z=-5$ visible as red stripes in the Fig.~\ref{fig:uxuzuyquiver} (a)---, there is almost no discernible signal on the $u_z$ and $u_y$ fields that quantify the rolls intensity. The main observation of roll activity evidenced by the largest amplitudes recorded on the $u_z$ and $u_y$ fields occurs at the location where the corresponding streak in the $u_x$ field has a meandering shape, i.e. where the streak is wavy ---see the zone indicated by dashed lines in Fig.~\ref{fig:uxuzuyquiver} (a), (b) and (c). This first qualitative observation is in agreement with the mechanism of roll regeneration, in which wavy streaks induce rolls \cite{Waleffe1997SSP}. In the following of the article, we describe quantitative results about this process.

A first quantification can be made looking at the spatial length scales related to the streaks and rolls: a horizontal profile of the streamwise velocity $u_x$ is shown in Fig.~\ref{fig:uxuzuyquiver} (d), corresponding to the green line parallel to the mean streaks in Fig.~\ref{fig:uxuzuyquiver} (a). The variation results from the streak undulation or waviness. The distance between two maximal fluctuations is the wavy wavelength $\lambda_{x \rm wavy}$.
A similar profile of the spanwise velocity $u_z$ is displayed in Fig.~\ref{fig:uxuzuyquiver} (e) in the middle of the undulation. This spanwise velocity is modulated along the wavy streak, and changes sign approximatively at the middle of the streak.  The distance between these two maximal fluctuations is half of the wavy wavelength $\lambda_{x \rm wavy}$. A vertical profile of the wall-normal velocity $u_y$ is shown in Fig.~\ref{fig:uxuzuyquiver} (f), corresponding to the  line at the $x$ position with zero $u_z$ velocity in Fig.~\ref{fig:uxuzuyquiver} (c). This spanwise velocity modulation is perpendicular to the wavy streak, and changes sign following the distance between streaks ${\lambda_z}/2$.

 
\begin{figure}[t]
	\centering
	\includegraphics[width=0.8\linewidth]{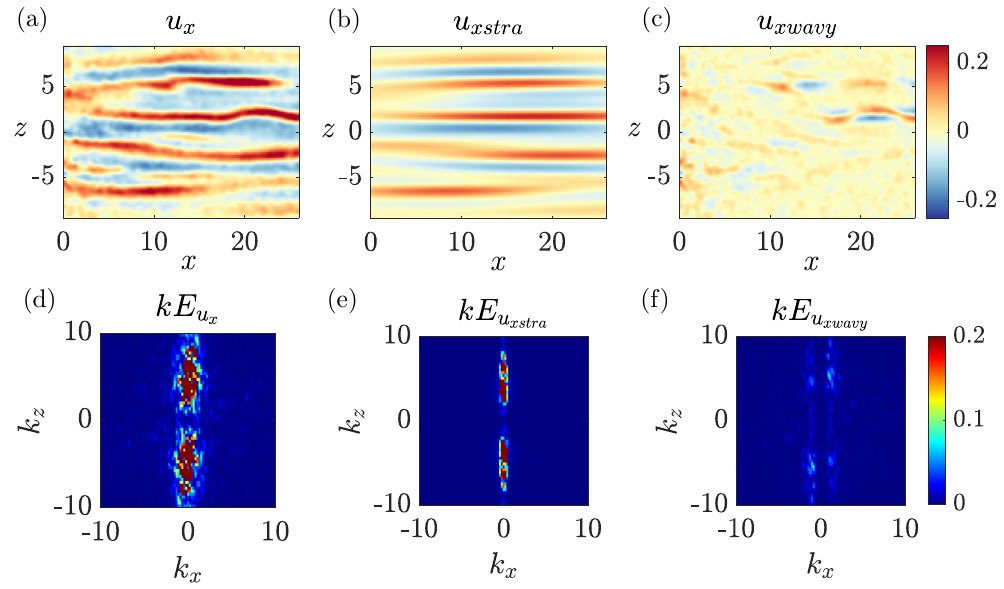}
	\caption{Decomposition, at $Re= 550$, of the small-scale streamwise velocity into straight and wavy components $u_x=u_{x\rm stra}+u_{x\rm wavy}$, and the corresponding pre-multiplied energy spectra.  ($a$) Small-scale flow $u_{x}$; ($b$) straight streaks $u_{x\rm stra}$; ($c$) wavy streaks $u_{x\rm wavy}$; ($d$) pre-multiplied energy spectrum $kE_{u_x}$ of $u_x$; ($e$) pre-multiplied energy spectrum $kE_{u_{x\rm stra}}$ of $u_{x\rm stra}$; ($f$) pre-multiplied energy spectrum $kE_{u_{x\rm wavy}}$ of $u_{x\rm wavy}$. }
	\label{fig:uxWAVYSTRA}
\end{figure} 

\subsection{Definition of streaks waviness}
\label{Definition of streak waviness}

In the example of Fig.~\ref{fig:uxuzuyquiver}, the spanwise and wall-normal velocities are maximum close to the undulating part of the streak. To investigate this link systematically, we introduce the ``wavy wall-normal vorticity' as a quantitative measurement of waviness. 
Using a Fourier filtering in the $x$ direction with a cut-off wavelength $\lambda_{x\rm wavy} \leq 14.5$  (or $k_{x\rm wavy}\geq {2\pi}/14.5$), we write the streamwise velocity $u_x$ as the sum of a straight component $u_{x\rm stra}$ and a wavy component $u_{x\rm wavy}$:
\begin{equation}
u_x=u_{x\rm stra}+u_{x\rm wavy} \,.
\end{equation} 
An example of this decomposition is shown in Fig.~\ref{fig:uxWAVYSTRA}.
The small scale velocity $u_x$ is shown in Fig.~\ref{fig:uxWAVYSTRA} (a) and is equal to the sum of the straight part $u_{x\rm stra}$  and the wavy part $u_{x\rm wavy}$, shown in Fig.~\ref{fig:uxWAVYSTRA} (b) and (c), respectively. With the chosen threshold value for $\lambda_{x\rm wavy}$, $u_{x\rm stra}$ actually appears straight, while the fluctuations of $u_{x\rm wavy}$ register the meandering of the streak. The corresponding pre-multiplied energy spectra used for this decomposition are shown in figures~\ref{fig:uxWAVYSTRA}(d), (e), and (f). 

With this decomposition, we can define the straight mode $u_{x\rm stra}$ as the zero mode of undisturbed streaks.
For Reynolds numbers below $Re_g$,  this mode exists because it is imposed by the vortex generators and is therefore independent of the turbulence level. 

Now, the straight streaks have a non-vanishing wall-normal vorticity $\omega_{y}$ due to the modulation of the velocity field in the $y$ direction. To measure the vorticity associated to the waviness, we define in the following Eq.~(\ref{eq:wavywallnormalvorticity}) a  ``wavy wall-normal vorticity'' denoted by $\omega_{y\rm wavy}$ based on $u_{x\rm wavy}$:
\begin{equation}
\omega_{y\rm wavy}= \frac{\partial u_{x\rm wavy}}{\partial z} - \frac{\partial u_{z}}{\partial x}
\label{eq:wavywallnormalvorticity}
\end{equation}
This ``wavy wall-normal vorticity'' $\omega_{y\rm wavy}$ vanishes for straight streaks but not for wavy streaks, and is larger when the streak is more undulated.

\begin{figure}[t]
	\centering
	\includegraphics[width=0.8\linewidth]{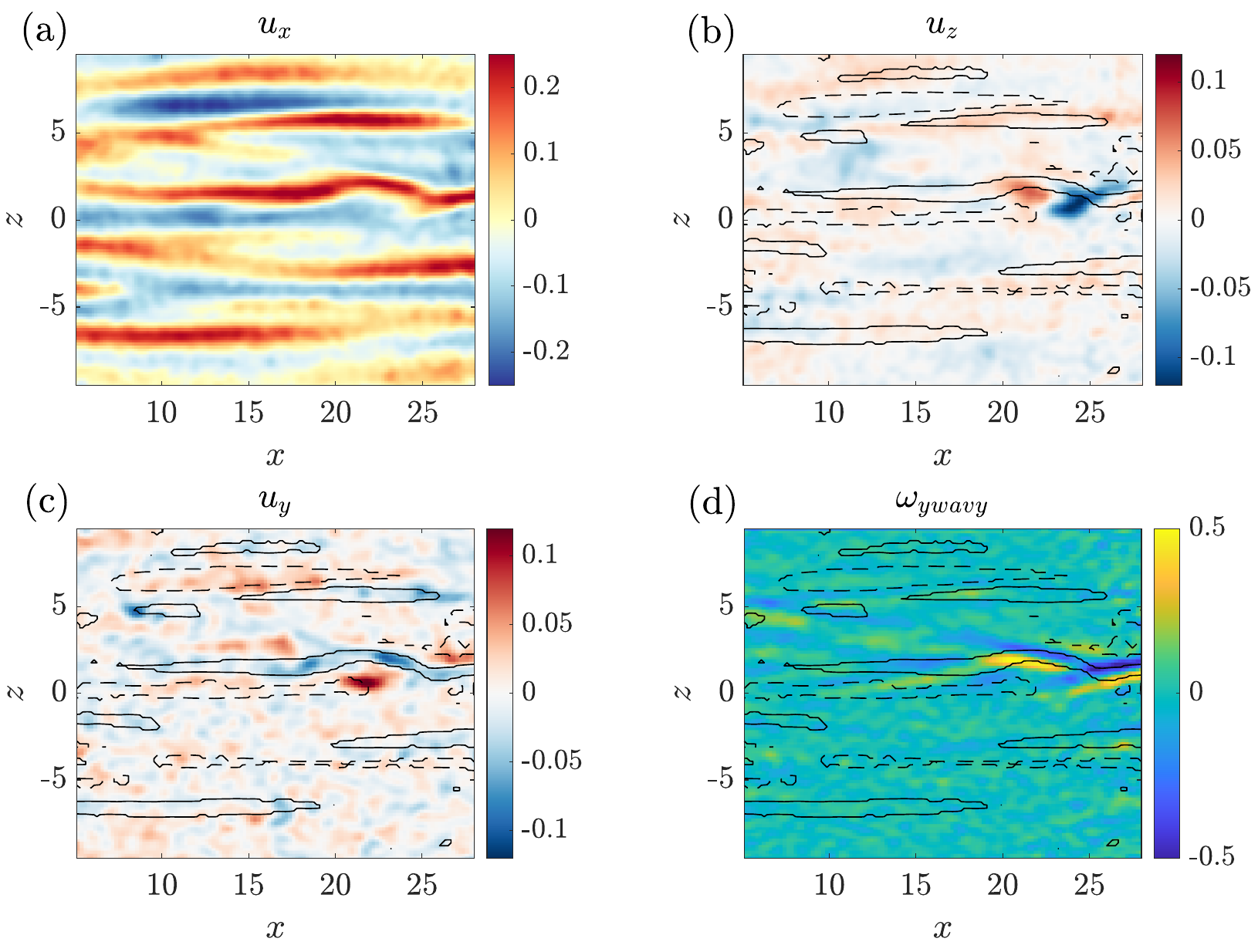}	
	\caption{Small scale flow snapshots at $Re=550$; (a) $u_x$ field; (b) $u_z$ field; (c) $u_y$ field; (d) $\omega_{y\rm wavy}$. On snapshots of $u_z$, $u_y$, $\omega_{y\rm wavy}$ the contour of the selected streaks are shown in solid lines ($u_x=0.1$) and dashed line ($u_x =-0.1$). See Supplementary Information for a video corresponding to the time evolution of this figure in a full experimental run.}
	\label{fig:wavystreaksrollsevolution}
\end{figure} 

\subsection{Local streaks detection and averaging}
\label{sec:Local streaks detection and averaging}

As discussed in the introduction, the role of waviness in the self-sustaining mechanism that regenerates turbulent structures is well established, although few laboratory experiments have investigated wavy streaks, like \cite{Mans2007} in boundary layers. In order to highlight this process, we investigate the relationship between streak waviness and the magnitude of velocity fluctuations from a local perspective.
We isolate the strongest streaks from the velocity field by thresholding to find the high speed (positive $u_x$) and low speed (negative $u_x$) regions, separately. An exemple of the streaks identified with such thresholding is shown in Fig.~\ref{fig:wavystreaksrollsevolution}, where positive and negative streaks are identified, respectively, by solid and dashed contours over the $u_z$, $u_y$, and $\omega_{y \rm wavy}$ fields. The chosen threshold on $|u_x|$ appears suitable to delimit streaks that enclose most of the perturbation on the other velocity components. A single dimensionless threshold $|u_x|>0.11$ is used for all Reynolds numbers tested. We define a ``streak-average'', denoted by $\left<|.|\right>$, as the average of the absolute value of the variable within the inner region of a streak. For example, $\left<|\omega_{y\rm wavy}|\right>$ for a given streak is the average of $|\omega_{y\rm wavy}|$ inside this streak. This value is slightly lower than the streak maximum.  Since the flow is perturbed by the presence of the vortex generators in the very neighborhood of their position at $x=0$, we only analyze the velocity field at positions with $x>5h$, where we can observe the natural streaks in the CPF.

\section{Results}

\subsection{Quantifying how streak waviness is related to roll amplitude}

\begin{figure}[t]
	\centering
	\includegraphics[width=0.8\linewidth]{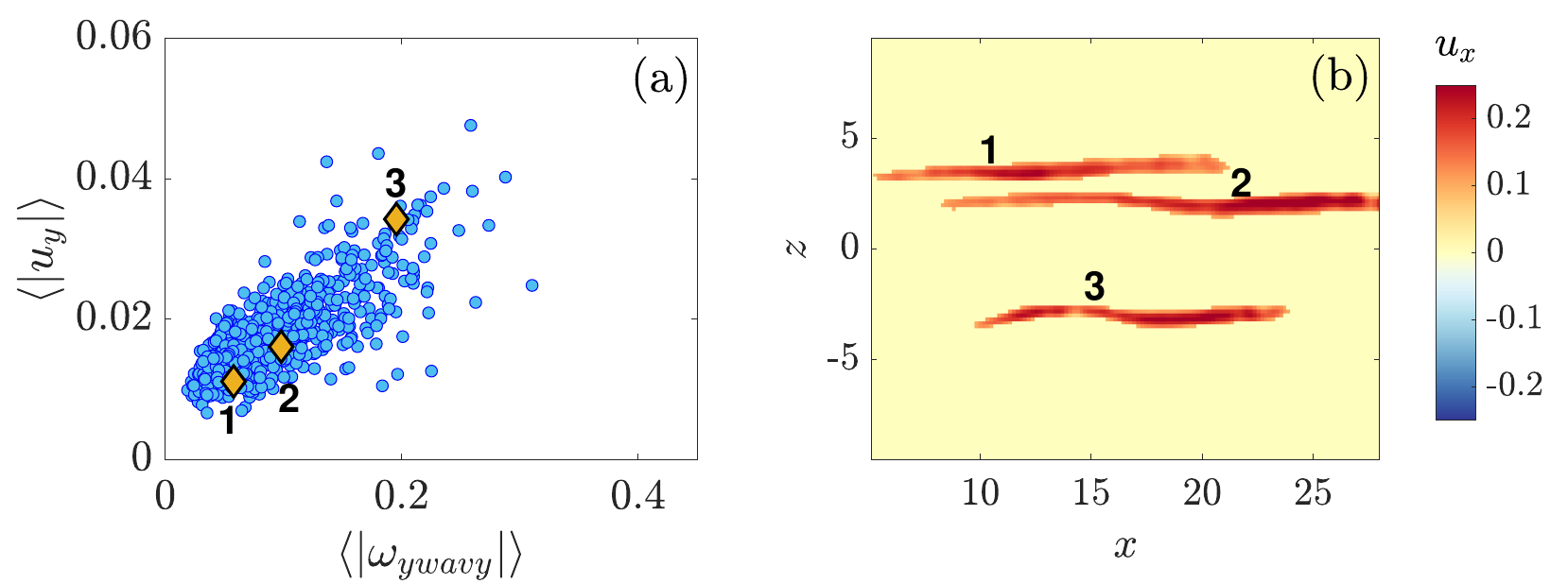}
	\caption{(a) Streak-averaged wall-normal velocity $\left<|u_{y}|\right>$ as a function of the wavy vorticity  $\left<|\omega_{y\rm wavy}|\right>$ for positive $u_x$ streaks at $Re=550$. (b) Streaks $u_x$ from one frame with numbers corresponding to the data points indicated in (a).}
	\label{fig:streaksshapeIllustration}
\end{figure}

\begin{figure}[thbp]
\centering 
\includegraphics[width=\linewidth,valign=t]{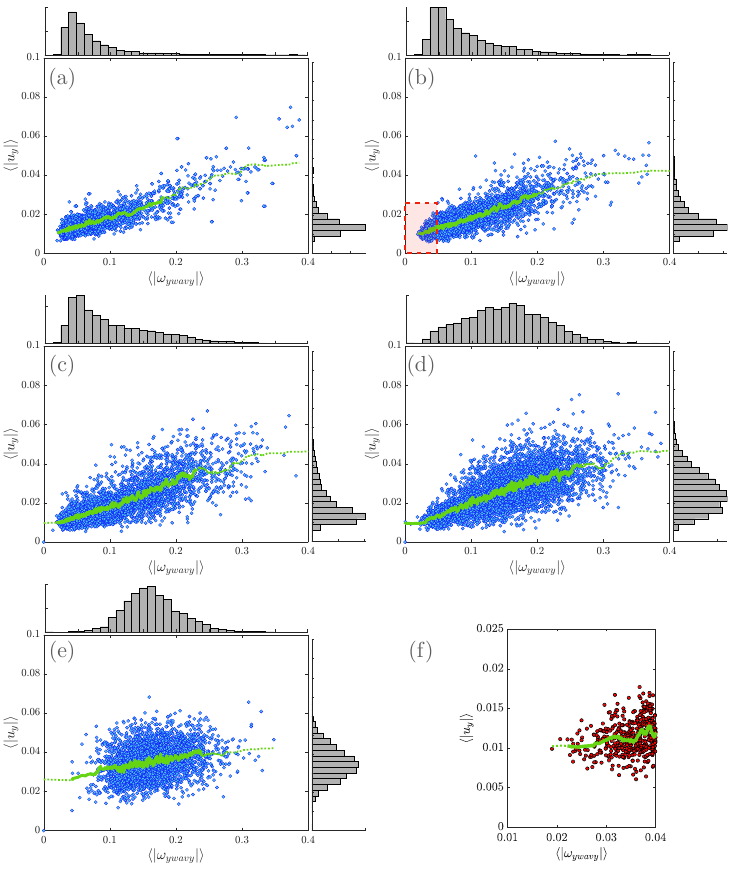}
\caption{Streak-averaged wall-normal $\left<|u_{y}|\right>$ velocity as a function of the streak-averaged wavy vorticity $\left<|\omega_{y\rm wavy}|\right>$ for (a) $Re=450$, (b) $Re=550$, (c) $Re=600$, (d) $Re=700$, and (e) $Re=1000$. The histograms on the top and right sides of each plot indicate the density of points at the corresponding position. (f) Zoom of the low-waviness region indicated in (b) showing that for the points with $\left<|\omega_{y\rm wavy}|\right> <0.04$ the increasing trend of $\left<|u_{y}|\right>$ as a function of$\left<|\omega_{y\rm wavy}|\right>$ cannot be established. On each curve a running average over a window of 50 points is shown as a solid green line; the line is broken when the number of points is scarce and the running average is no longer meaningful.}
	\label{fig:uy_omega_ywavy_Re}
\end{figure}

In order to gain statistical robustness on the qualitative observation given by Fig.~\ref{fig:wavystreaksrollsevolution}, we proceed to analyse the streak-averaged variables for all the experiments performed. We use in the following analysis the high-speed streaks. A quantitative picture of the link between the strength of the rolls and the waviness of the streaks is given in Fig.~\ref{fig:streaksshapeIllustration} by plotting the wall-normal streak-averaged velocity $\left<|u_{y}|\right>$  as a function of the wavy vorticity $\left<|\omega_{y\rm wavy}|\right>$ for the positive streaks of one experiment at $Re=550$. One experiment means the whole measurement sequence represented in the Supplementary Video corresponding to Fig.~\ref{fig:wavystreaksrollsevolution}, where several streaks can be identified on each recorded frame. Each data point in Fig.~\ref{fig:streaksshapeIllustration} (a) corresponds to a local averaging on one streak at a given time. We process only one frame every five, which is approximately every $1.1$ advection times $h/U_{\rm belt}$, to avoid obtaining several data points from that would be associated to the same streaks during the temporal evolution. The main observation appears clearly: higher wall-normal velocity fluctuations $\left<|u_{y}|\right>$ are associated to higher waviness of the streaks $\left<|\omega_{y\rm wavy}|\right>$. The same qualitative observation is obtained when examining the spanwise velocity fluctuation $\left<|u_{z}|\right>$ as a proxy for the roll strength (not shown here).
 
We show the shape of streaks with different waviness in Fig.~\ref{fig:streaksshapeIllustration} ($b$), where the numbers correspond to the points labeled in Fig.~\ref{fig:streaksshapeIllustration} ($a$). Streak $1$, with a nearly flat elongated shape, is associated to the very weak but non-zero amplitude of the roll variable $\left<|u_{y}|\right>$. Streak $2$, with a slightly meandering shape, corresponds to higher values of $\left<|u_{y}|\right>$. This tendency becomes more pronounced for streak $3$, which has a stronger undulation and corresponds to even higher amplitude rolls, compared to streak $2$. 

In Fig.~\ref{fig:uy_omega_ywavy_Re}, the streak-averaged wall-normal  $\left<|u_{y}|\right>$ velocity amplitudes are plotted as a function of the wavy wall-normal vorticity $\left<|\omega_{y\rm wavy}|\right>$ for all the Reynolds numbers tested. The plots are thus the equivalent to Fig.~\ref{fig:streaksshapeIllustration} (a), but for each Reynolds number 5 independent experimental runs are analyzed. This gives between 3000 and 5000 points for each Reynolds number, which allows us to get very good statistics. The measurements were conducted on positive streaks due to their higher instability, whereas negative streaks are found to be more unstable in wall turbulence \cite{MansPhdThesis}. This difference can be explained by the base profile of the Couette-Poiseuille flow in turbulence, which features a high shear region on the moving wall and an almost flat region on the opposite side caused by back-flow.

It can be seen in Fig.~\ref{fig:uy_omega_ywavy_Re} that the number of data points with higher waviness grows with the Reynolds number. A running average is superimposed on each plot. Recalling that the  wall-normal velocity $\left<|u_{y}|\right>$ quantifies the roll intensity, the increasing value of their running average as a function of $\left<|\omega_{y\rm wavy}|\right>$, quantitatively confirms their relationship to the streak waviness, as expected from models describing the Self-Sustained Process (SSP) to regenerate turbulence \cite{Waleffe1997SSP}. In fact, even for a Reynolds number lower than the critical value $Re_g$, the streaks generated by the vortex generators with a strong undulation show a vigorous activity.  As observed in the histograms, the most probable value of $\left<|\omega_{y\rm wavy}|\right>$ increases dramatically when the Reynolds number changes from $600$ to $700$, i.e. around the self-sustained threshold $Re_g\approx680$. This is a supplementary indication that waviness plays a major role in the self-sustained mechanism.

\begin{figure}[t]
	\centering
\includegraphics[width=0.65\linewidth]{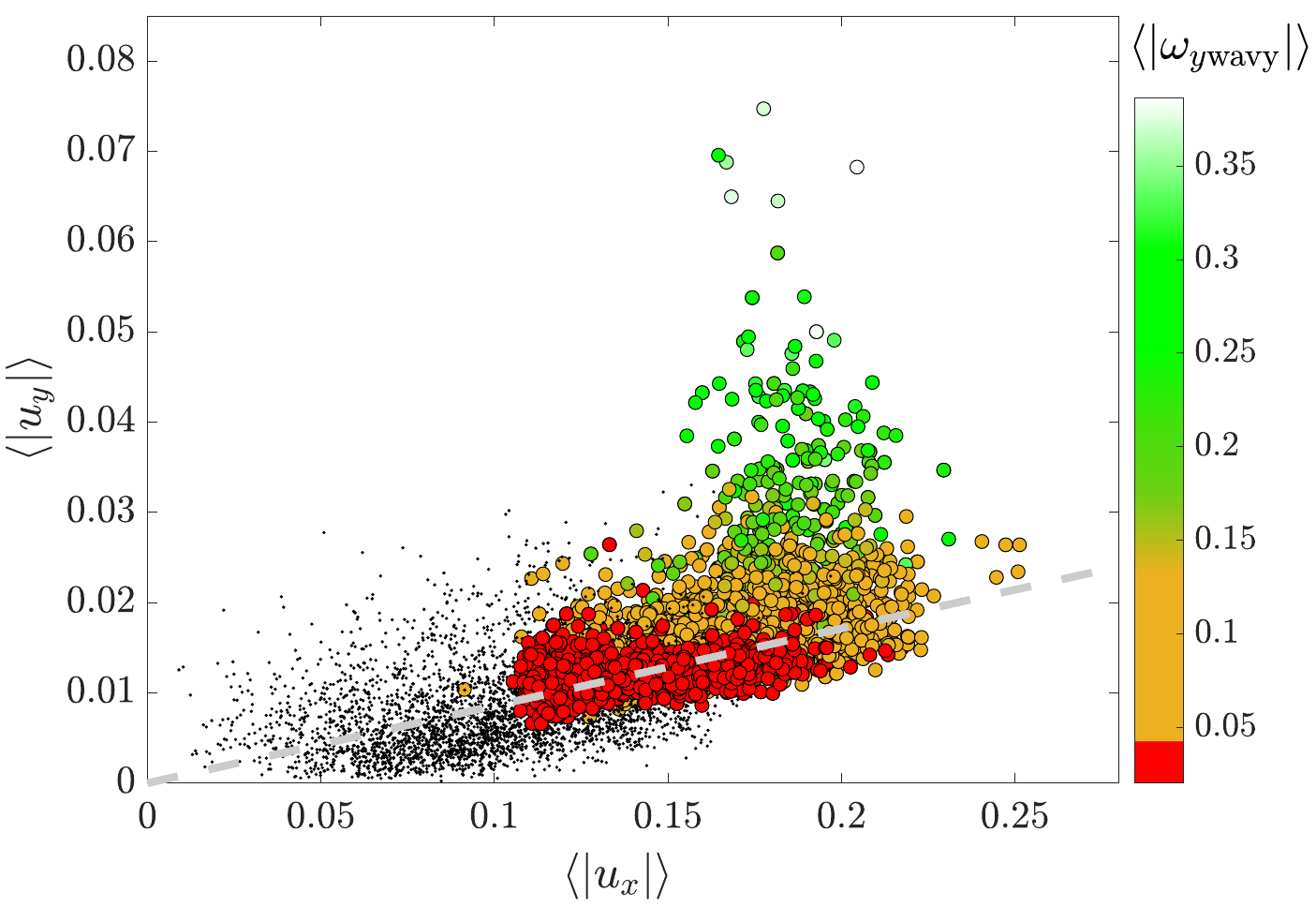}
	\caption{$\left<|u_{y}|\right>$ vs.  $\left<u_{x}\right>$ for all 5 experiments at $Re=550$ (the colormap indicates values of $\left<|\omega_{y\rm wavy}|\right> $). The red circles correspond to cases values of $\left<|\omega_{y\rm wavy}|\right> < 0.04$ (a linear fit of these values is shown in dashed line). The black dots are the filtered fields $\left<|u_{y \rm stra}|\right>$ vs. $\left<|u_{x \rm stra}|\right>$. A linear fit of the black points is indistinguishable within the accuracy of the experimental measurements from the dashed line. }
	\label{fig:uy_ux_all}
\end{figure}


\subsection{Lift-up characterisation and SSP process} 
\label{Lift-up characterisation and SSP process}

In this section, we investigate the link between the roll amplitude variable $\left<|u_{y}|\right>$ and the streamwise velocity $\left<|u_{x}|\right>$ that characterises the streak intensity. We are thus looking into the basic mechanism engendered by streaks in wall-bounded flows: the lift-up effect \cite{Brandt:2014_LiftupEffectLinear,lozano-duran2021,Jiao2021}.  

In order to examine the lift-up, we focus on streaks with low waviness. For these cases $\left<|\omega_{y\rm wavy}|\right>$ is no longer a good variable, as shown in Fig.~\ref{fig:uy_omega_ywavy_Re}~(f) that presents a magnified view of figure~\ref{fig:uy_omega_ywavy_Re}~(b) for $\left<|\omega_{y\rm wavy}|\right> < 0.04$. In this restricted range of small $\left<|\omega_{y\rm wavy}|\right>$, the mean value of $\left<|u_{y}|\right>$ is almost constant, and the dispersion of the points is large. The same points are plotted in red in Fig.~\ref{fig:uy_ux_all} that shows the wall-normal velocity $\left<|u_{y}|\right>$ as a function of the streamwise velocity  $\left<|u_{x}|\right>$.  These points are well-described by a straight line  $\left<|u_{y}|\right> \approx 0.08 \left<|u_{x}|\right>$ (dashed line in Fig.~\ref{fig:uy_ux_all}). 
In this case of absence of meandering, the existence of straight streaks is a manifestation of the ``laminar'' lift-up mechanism, since the presence of rolls built up by the vortex generators modulates the streamwise velocity and produces high and low velocity streaks.

Another more precise approach to quantify this ``laminar'' lift-up  is to use the decomposition of all the streaks, including the very wavy ones, as the sum of straight and wavy components introduced earlier (see section~\ref{Definition of streak waviness}). The value of $\left<|u_{y \rm stra}|\right>$ as a function of  $\left<|u_{x \rm stra}|\right>$ is shown as black dots in Fig.~\ref{fig:uy_ux_all}.  These are more dispersed and reach lower values of $\left<|u_{y}|\right>$ and $\left<|u_{x}|\right>$  than the low-waviness points (the red points in Fig.~\ref{fig:uy_ux_all} that correspond to $\left<|\omega_{y\rm wavy}|\right> < 0.04$). However, a linear fit of these black points is in practice almost identical to the fit of the red points shown as a dashed line in Fig.~\ref{fig:uy_ux_all}, indicating a similar physical mechanism.

\begin{figure}[hp!]
	\centering
\includegraphics[width=0.95\linewidth,valign=t]{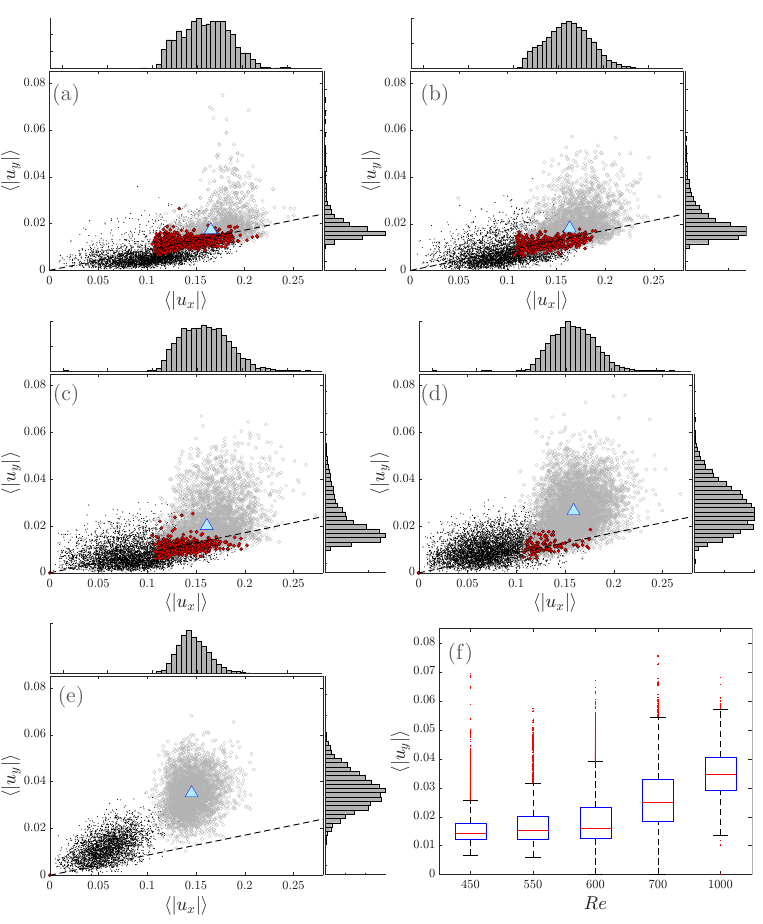}
	\caption{$\left<|u_{y}|\right>$ vs.  $\left<|u_{x}|\right>$ for all Reynolds numbers: (a) $Re=450$, (b) $Re=550$, (c) $Re=600$, (d) $Re=700$ and (e) $Re=1000$. The histograms on the top and right sides of each plot indicate the density of points at the corresponding position. The mean values are shown as a blue triangle on each plot. The points where $\left<|\omega_{y\rm wavy}|\right> < 0.04$ are marked in red and the black dots are the filtered fields $\left<|u_{y \rm stra}|\right>$ vs. $\left<|u_{x \rm stra}|\right>$, as in Fig.~\ref{fig:uy_ux_all}. The fit for the red points at $Re=450$ is recalled as a dashed line in all plots, showing that it serves as a lower bound and that in the turbulent regimes all points are above it.  (f) Box plots representing the histogram of $\left<|u_{y}|\right>$ for each Reynolds number. On each box, the central red line indicates the median, and the bottom and top edges of the box indicate the 25th and 75th percentiles, respectively. The whiskers extend to the most extreme data points not considered outliers, and the outliers are the red markers. }
	\label{fig:uy_ux_Re}
\end{figure}

In Fig.~\ref{fig:uy_ux_Re} the relation between the rolls variable $\left<|u_{y}|\right>$ and streaks variable $\left<|u_{x}|\right>$ is displayed for different Reynolds numbers: $Re= 450, 550, 600, 700$ and $1000$. For all these Reynolds numbers, the low-waviness points (red markers in Fig.~\ref{fig:uy_ux_Re}) are located around the straight line $\left<|u_{y}|\right> = 0.08 \left<|u_{x}|\right>$. These red points correspond to the lower bound of the cloud of points, which is almost unchanged for the three more laminar cases ($Re= 450, 550$ and $600$).

At higher Reynolds numbers the points are more scattered. At $Re=700$, only a few red points remain close to the straight line, and at $Re=1000$, due to the fully sustained turbulent nature of the flow at this Reynolds number, there are no more red points: the cloud of (gray) points is located well above the straight line, outside the region of weak undulation. This is consistent with a fully self-sustained dynamics at $Re=1000$ and with the $\left<|u_{y}|\right>$ histogram on the figures, which shows that the higher the Reynolds number, the higher the average value of the rolls amplitude $\left<|u_{y}|\right>$ (marked by the blue triangles in Fig.~\ref{fig:uy_ux_Re}). When all the streaks are wavy, $\left<|u_{y}|\right>$ is larger, in agreement with the spatially organized process of near-wall structure regeneration for self-sustained turbulence (SSP) formulated by Waleffe \cite{Waleffe1997SSP}. 

It is also interesting to note that if one were to attempt a linear fit of the black points that represent $u_{y \rm stra}$ vs. $u_{x \rm stra}$, its slope would increase strongly for $Re=1000$, with respect to the laminar lift-up slope represented by the dashed line in Fig-~\ref{fig:uy_ux_Re}. This is a signature that when the flow is fully turbulent, nonlinear effects dominate. This is because the lift-up phenomenon depends on the gradient of the base velocity profile  \cite{Landahl1975}, and the base velocity profile is nonlinearly modified by the mean flow distortion.

\section{Conclusions }

We have investigated experimentally the dynamics of turbulent structures, streaks and rolls, in a Couette-Poiseuille channel. The interaction between streaks and rolls has been shown to be a useful tool to interpret the lift-up and the self-sustaining process of wall-bounded turbulence around a threshold Reynolds number. This work provides an experimental evidence of certain steps in the SSP models and proposes a new way to quantify the relation between waviness and amplitude of the streaks, which has not been shown in previous works. 

One of the main results of the present article is the experimental quantification of how the roll intensity increases with streak waviness. Waviness is a key variable in the roll regeneration equation of SSP reduced models ---see the third equation of the system (20) in \cite{Waleffe1997SSP}. To quantify the waviness, we have defined a ``wavy wall-normal vorticity'' by high-pass filtering $u_x$, i.e. by removing the straight component of the streaks.   By considering the average values of the waviness and of the velocity component on a single streak at a given time, we have determined a relationship that is local in space and time.  We have shown that for the investigated Reynolds numbers between $450$ and $1000$ the average intensity of the rolls increases when the wavy vorticity increases. We observed in particular that the structures that do not contribute to the turbulence regeneration are those that we have defined as straight streaks, confirming the claim in the numerical work of Jimenez \cite{Jimenez:2022} that these streaks do not participate in wall turbulence.

Another key result is that for low-waviness streaks, the average intensity of the rolls $\left<|u_{y}|\right>$ is proportional to $\left<|u_{x}|\right>$, with $\left<|u_{y}|\right> \approx 0.08 \left<|u_{x}|\right>$. Remarkably, $\left<|u_{y}|\right>$ as a function $\left<|u_{x}|\right>$ of the straight part of the streaks is shown to align with the same slope as the one obtained for the low-waviness streaks. This behavior is observed even if the original streaks (i.e. before low-pass filtering) are very wavy, and can be interpreted a as a result of a laminar lift-up. This is consistent with the second equation of  system (20) in the reduced model of Waleffe \cite{Waleffe1997SSP}, of equation (22) of the model of Moehlis \cite{Moehlis:2004} and equations (9d) and (9k) of the model of Cavalieri \cite{Cavalieri:2021}, where a linear relationship between streak intensity and rolls is described.
However, a direct comparison with the coefficients of the reduced models cannot be performed, since these equations are derived for different boundary conditions and flow profiles in the wall-normal direction, and also include other terms such as a mean flow correction that are not available from our PIV measurements in a plane.

Some of the results of the present article are specific to the Couette-Poiseuille flow. For instance, we have analysed the positive streaks due to their higher instability, whereas negative streaks are found to be more unstable in boundary layer turbulence. On the other hand, the link between rolls intensity and waviness, or between rolls intensity and streak intensity for low-waviness streaks, is expected to be also verified for other confined shear flows like Couette or Poiseuille flows, and even in flows showing supercritical instabilities with streamwise structures as in centrifugal instabilities. The physical mechanisms of the self-sustaining of turbulence are thought to be the same in these geometries. Waviness has been shown to be a key parameter in wall turbulence, where larger structures display a larger mean level of waviness \cite{Hwang:2022}. 

\bigskip
We acknowledge Amaury Fourgeaud, Olivier Brouard, Xavier Benoit-Gonin, and Thierry Darnige, from the PMMH technical support for their patient and invaluable help with the experimental setup.

\providecommand{\noopsort}[1]{}\providecommand{\singleletter}[1]{#1}%

\end{document}